\documentclass[10pt,a4paper,notitlepage]{revtex4}
\pdfoutput=1
\usepackage{amsmath,amssymb,graphicx}
\usepackage{color}
\usepackage{subfigure}
\usepackage{titlesec}
\usepackage{indentfirst}
\begin{document}
\fontsize{10}{13}
\selectfont
\titlespacing*{\section}
{0pt}{2ex plus 1ex minus .2ex}{1.5ex plus .2ex}

\title{Complex Doppler effect in left-handed metamaterials}
\author{D. Ziemkiewicz}
\email{david.ziemkiewicz@utp.edu.pl}
\author{S. Zieli\'nska-Raczy\'nska}
\affiliation{Institute of Mathematics and Physics, University of Technology and Life Sciences, Al. Kaliskiego 7, 85-798 Bydgoszcz, Poland.}
\begin{abstract}
The Doppler shift is investigated in one-dimensional system with moving source. Theoretical findings are confirmed by numerical simulations of optical and acoustical waves propagation in a simple metamaterial model, showing the reversed shift and the existence of multiple frequency modes. The properties of these waves are discussed and the effect of absorption on the phenomenon is outlined.
\end{abstract}
\maketitle
\section{Introduction}
Veselago's pioneering paper on media which have negative index of optical refraction \cite{Veselago} followed by practical demonstration in metamaterials \cite{Smith,Shelby} initiated a time of intensive studies on their properties, fabrication and applications \cite{Veselago2,Pendry1,Liu}. Such media, commonly called left-handed materials (LHM), referring to the fact that $\vec E$, $\vec H$ and $\vec k$ vectors of electromagnetic wave propagating inside them forms a left-handed system, exhibit many peculiar properties such as negative refraction \cite{Veselago,Shelby}, reversed Goos - H\"{a}nchen shift \cite{GHshift}, evanescent wave amplification and diffraction limit breaking \cite{Pendry2} just to name a few. The reversed Doppler effect, described by  Veselago \cite{Veselago}, is one of the most characteristic features of negative index media. In these systems, in opposition to the right-handed materials, the frequency of the signal emitted by a moving source approaching the observer decreases and the signal from the receding source is upshifted. The phenomenon, which occurs for electromagnetic as well as sound waves, has been observed in a wide range of systems, including photonic crystals \cite{Dop_Opt}, magnetic thin film \cite{Spinwave}, transmission lines \cite{Seddon} and acoustical metamaterials \cite{Sound_Lee}. Another type of effects which leads to the frequency shift is so-called rotational Doppler effect. This phenomenon occurs when light  beam is rotated around its propagation axis and is associated with light beam carrying orbital angular momentum \cite{Luo}.

Our paper is devoted to the study  the Doppler effect in a one-dimensional system consisting of dispersive left-handed medium and infinitely small source of plane waves. Assuming simple Drude model of material dispersion, the frequency and wavelength shifts are discussed. The so-called complex Doppler effect \cite{Frank} resulting in generation of multiple frequency modes is investigated in the case of negative refraction index. Here we give relatively simple analytic formulas which allow one to qualitatively indicate the region of experimental parameters for which these frequency modes could be observed. The two distinct sets of solutions are applicable in the optical regime at up to relativistic source velocities and in the acoustical regime where the considered speeds are more easily realizable. The complex Doppler effect has been studied by a number of authors; Bazhanova \cite{Bazhanova} considered the effect in plasma and Lisenkov \textit{et al} \cite{Lisenkov} analyzed the phenomenon in double negative media by using Green's function formalism to solve Helmholtz equation. 
Here we have used first principles approach and confirmed the validity of our theoretical predictions numerically using the Finite Difference Time Domain (FDTD) method, which is one of the most flexible algorithm commonly used to study wave propagation in metamaterials \cite{Veselago2,KS_Taflove,Ziolkowski,FDTD_Lee}. The performed numerical simulations deal with the problem of a source moving inside negative index medium, as opposed to the more extensively studied case of reflection from moving metamaterial interface \cite{Xiao,Chiou} or other systems where additional frequency components might be present due to the metamaterial slab surface modes \cite{Wang}. In the case of acoustical system, the complex Doppler effect was examined for both left- and right-handed media, showing remarkable difference between these cases.
\section{Doppler effect in one-dimensional system}
\begin{figure}[ht!]
    \centering
    \includegraphics[scale=0.2]{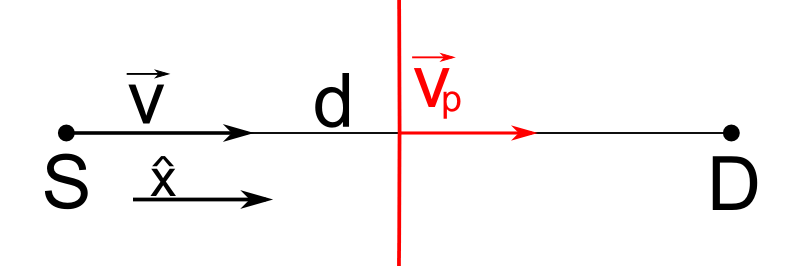}
    \caption{Wavefront sent by a moving source S towards detector D.}
    \label{fig:1}
\end{figure}

Consider a point radiation source S located at $x=0$, emitting waves towards detector $D$ at $x=d$. The two consecutive wavefronts are emitted at $t_1 = 0$ and $t_2 = T_0$ respectively, where $T_0=\frac{2\pi}{\omega_0}$ is the period of the source. In the reference frame of the detector, the source is moving at a constant velocity $\vec V=V \hat x$ and the medium is stationary. The system, as seen in the frame $D$, is shown on the Fig. \ref{fig:1}. 
\newpage\noindent
In the reference frame of the detector, the wave will have some shifted frequency $\omega^+$. Assuming that the medium is dispersive, the wave fronts will propagate at a frequency dependent phase velocity $V_p(\omega^+)$. Two consecutive fronts will reach the detector at 
\begin{eqnarray}\label{System1}
t_1 &=& \frac{d}{v_p(\omega^+)}=\frac{n(\omega^+)d}{c},\nonumber\\
t_2 &=& \frac{T_0}{\sqrt{1-\beta^2}} + \frac{n(\omega^+)\left(d - V\frac{T_0}{\sqrt{1-\beta^2}}\right)}{c},
\end{eqnarray}
where $n(\omega^+)$ is the refraction index of the medium and $\beta = V/c$. The period of the detected wave is
\begin{equation} \label{Dop1}
T^+ = t_2-t_1 = T_0\frac{1-n(\omega^+)\beta}{\sqrt{1-\beta^2}}.
\end{equation}
The above result can be also obtained in a straightforward way from the relativistic phase invariance which yields
\begin{eqnarray} \label{transf1}
\omega_0 = \frac{\omega^+ - vk^+}{\sqrt{1-\beta^2}},&\quad& \omega^+ = \frac{\omega_0 + vk_0}{\sqrt{1-\beta^2}}, \nonumber\\
k_0 = \frac{k^+ - \omega^+ \frac{v}{c^2}}{\sqrt{1-\beta^2}}, &\quad& k^+ = \frac{k_0 + \omega_0 \frac{v}{c^2}}{\sqrt{1-\beta^2}},
\end{eqnarray}
where $k_0 = n_0\omega_0/c$ is the wave vector in the frame of the source and $n_0$ is the refraction index of the moving medium given by \cite{Grzegorczyk}
\begin{equation} \label{nprim}
n_0=\frac{n(\omega^+) - \beta}{1 - n(\omega^+)\beta}.
\end{equation}
This result can be obtained from relativistic velocity addition. In the case when source speed $V$ exceeds the phase speed $V_p$, we have $n(\omega^+)\beta > 1$ and, in consequence, as seen in the Eq. \ref{System1}, the wave emitted later will arrive at the detector first. As the period is a positive quantity, the final equation for Doppler shift for source approaching (+) and receding from the detector (-) is given by implicit relation
\begin{equation} \label{Dop3}
\omega^\pm = \omega_0\left|\frac{\sqrt{1-\beta^2}}{1\mp n(\omega^\pm)\beta}\right|.
\end{equation}
Multiple solutions of Eq. \ref{Dop3} correspond to the complex Doppler effect \cite{Bazhanova,Lisenkov}. In such a case, a monochromatic source of frequency $\omega_0$ generates a number of wave modes with detected frequencies $\omega_i$ and their associated wavevectors $k_i = \omega_i n(\omega_i)/c$. Interestingly, the transformation given by Eq. \ref{nprim} is a one to one correspondence, so it has the property that different values of $n(\omega_i)$ cannot yield single value of $n_0$. This means that, in contrast to the derivation presented in \cite{Berger}, we can assume that in the reference frame of the source, the wave is also split into multiple modes of the same frequency $\omega_0$ but of different wavevectors $k_{0i}$. Only under such condition the transformations in Eq. \ref{transf1} remain consistent for multiple solutions.

The above derivation can be adapted to sound waves by skipping the relativistic term $\sqrt{1-\beta^2}$ and defining $c$ as a speed of sound in the air in respect to which all refraction indices are defined. In left-handed medium the value of $n(\omega_0)$ is negative; this causes the reversal of Doppler effect - the wave emitted in the direction of motion is downshifted ($\omega^+ < \omega_0$) and the backwards going one is upshifted
($\omega^- > \omega_0$). The wavelength measured in the detector frame is described by the relation
\begin{equation}\label{Lambda}
\lambda^\pm = \frac{2\pi c}{\left|n(\omega^\pm)\right|\omega^\pm}.
\end{equation}
Again, the absolute value is used to ensure that $\lambda^\pm$ is a positive quantity.

\section{Analytic solution for Drude model}
\begin{figure}[ht!]
    \centering
    \includegraphics[scale=0.3]{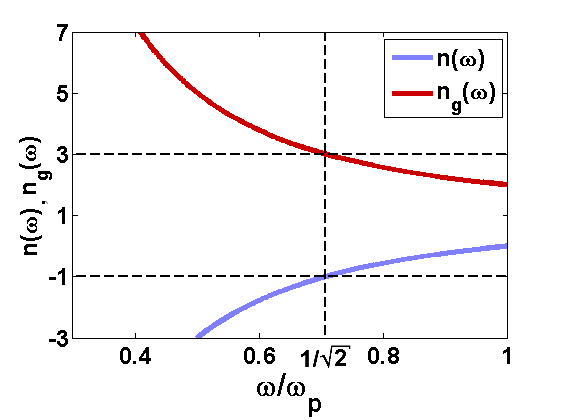}
    \caption{Phase and group refraction indices in a Drude model metamaterial.}
    \label{fig:2}
\end{figure}

One of the simplest dispersion relations for the dielectric permittivity $\epsilon$ and magnetic permeability $\mu$, considered in the Veselago's original work \cite{Veselago}, are given by the Drude model
    \begin{eqnarray} \label{Dr_Model}
    \epsilon(\omega)= 1 - \frac{\omega_{pe}^2}{\omega^2},\nonumber\\
    \mu(\omega)= 1 - \frac{\omega_{pm}^2}{\omega^2}.
    \end{eqnarray}
In metamaterials, the so-called plasma frequencies $\omega_{pe}$ and $\omega_{pm}$ are determined by the medium structure \cite{Pendry_8}. The Eq. \ref{Dr_Model} is an exact solution for magnetic permittivity of metamaterial composed of long, thin wires \cite{Pendry_8}. For magnetic permeability, the proposed relation is a good approximation of a full Drude-Lorentz model in the frequency range above resonance, where the permeability is negative \cite{Pendry_6}. In the simplest case, for $\omega_{pe} = \omega_{pm} = \omega_p$, the refraction index $n$ is described by
\begin{equation} \label{modDRU}
n(\omega) = 1 - \frac{\omega_p^2}{\omega^2}.
\end{equation}
Another important property of the medium is the group velocity defined as
\begin{equation}
v_g = \frac{\partial \omega}{\partial k} = \frac{c}{n_g}
\end{equation} 
where $n_g$ is the group refraction index is given by
\begin{equation}
n_g(\omega)=n(\omega)+\omega\frac{\partial n(\omega)}{\partial \omega} = 1 + \frac{\omega_p^2}{\omega^2}.
\end{equation}
The dispersion relation for such material is shown on the Fig. \ref{fig:2}. To further simplify the description of the Doppler effect, one can set the nominal source frequency to $\omega_0 = \frac{1}{\sqrt{2}}\omega_p$, which leads to the refraction index $n(\omega_0)=-1$ and, accordingly, a negative phase velocity $v_p = -c$, a characteristic feature of LHM.
\newline\noindent
By inserting the Eq. \ref{modDRU} into the Doppler shift Eq. \ref{Dop3}, one obtains
\begin{equation} \label{Dop5}
\omega^\pm = \omega_0\left|\frac{\sqrt{1-\beta^2}}{1\mp\beta \pm \beta\left(\frac{\omega_p}{\omega^\pm}\right)^2}\right|.
\end{equation}
The solutions of the Eq. \ref{Dop5} describe up to four frequency components
\begin{eqnarray} \label{Final}
\omega_1^\pm(\beta) = \frac{\sqrt{1-\beta^2}\omega_0+\sqrt{(1-\beta^2)\omega_0^2\mp 4\beta(1\mp \beta)\omega_p^2}}{2(1\mp\beta)},\nonumber\\[1em]
\omega_2^\pm(\beta) = \frac{\pm\sqrt{1-\beta^2}\omega_0\mp\sqrt{(1-\beta^2)\omega_0^2\mp 4\beta(1\mp \beta)\omega_p^2}}{2(1\mp\beta)}.
\end{eqnarray}
The waves of frequency $\omega_1^\pm$ are the usual Doppler modes that diverge from the value of $\omega_0$ as the velocity $V$ increases. The additional modes $\omega_2^\pm$ approach 0 as the $\beta \rightarrow 0$. For the forward going waves, the range of source velocity where the solutions are real valued is limited. In the border case of a single real solution, the group velocity of the generated wave is equal to the source velocity. 

The medium described by the dispersion Eq. \ref{modDRU} is a fair approximation of the acoustical metamaterial presented in \cite{Sound_Lee} in the region of negative refraction index. By dropping the relativistic term $\sqrt{1 - \beta^2}$ from the derivation of the Doppler effect, one obtains four solutions for shifted modes  
\begin{eqnarray} \label{Final_D}
\omega_1^\pm(\beta) = \frac{\omega_0+\sqrt{\omega_0^2\mp 4\beta(1\mp \beta)\omega_p^2}}{2(1\mp\beta)},\nonumber\\[1em]
\omega_2^\pm(\beta) = \frac{\pm\omega_0\mp\sqrt{\omega_0^2\mp 4\beta(1\mp \beta)\omega_p^2}}{2(1\mp\beta)},
\end{eqnarray}
where $\beta$ is the ratio of the source velocity $V$ to the speed of sound in the air.
\section{Simulation setup}
The simulation is based on a standard Yee formulation of FDTD algorithm \cite{Yee} realized in one dimension, describing a plane wave traveling along the $\hat x$ direction with electric and magnetic field vectors $\vec E = [0,E,0]$, $\vec H = [0,0,H]$. By using a normalization $\epsilon_0 = \mu_0 = c = 1$, from the Maxwell's equations one gets
\begin{eqnarray}
-\frac{\partial}{\partial x} E &=& j_m + \sigma_m H +\frac{\partial}{\partial t}\left(H + M\right),  \nonumber\\
-\frac{\partial}{\partial x} H &=& j_e + \sigma_e E + \frac{\partial}{\partial t}\left(E + P\right),
\end{eqnarray}
where $P$ and $M$ are the values of polarization and magnetization vectors in the medium, $\sigma_e$, $\sigma_m$ describe the electric and magnetic conductivity and $j_e$, $j_m$ are externally imposed current densities . The simulation space is divided into set of discrete spatial points with defined values of field components $E_y(x,t)$ and $H_z(x,t)$, separated by equal distance $\Delta x = 1$. The time is quantized by a time step $\Delta t = 0.5$, bounded by the stability of the dispersion model calculation scheme. By introducing the notation
\begin{equation}
E(x\Delta x, n\Delta t) = E_x^n,
\end{equation}
one can write the final, discretized equations describing field evolution
\begin{eqnarray}
H_{x+\frac{1}{2}}^{n+\frac{1}{2}} &=& \frac{2 - \sigma_m \Delta t}{2 + \sigma_m \Delta t} H_{x+\frac{1}{2}}^{n-\frac{1}{2}} - \frac{2\Delta t}{2 + \sigma_m \Delta t}\left[\frac{E_{x+1}^n - E_x^n}{\Delta x} + \frac{M_{x+\frac{1}{2}}^{n+\frac{1}{2}}-M_{x+\frac{1}{2}}^{n-\frac{1}{2}}}{\Delta t} +j_m\right], \nonumber\\
E_x^{n+1} &=& \frac{2 - \sigma_e \Delta t}{2 + \sigma_e \Delta t} E_x^{n} - \frac{2\Delta t}{2 + \sigma_e \Delta t}\left[\frac{H_{x+\frac{1}{2}}^{n+\frac{1}{2}} - H_{x-\frac{1}{2}}^{n+\frac{1}{2}}}{\Delta x} + \frac{P_{x}^{n+1}-P_{x}^{n}}{\Delta t} + j_e\right].
\end{eqnarray}\newpage\noindent
The waves are generated by monochromatic, electric current source $j_e(t)$ with normalized frequency $\omega=0.05$, corresponding to the period $T = 40\pi \Delta t$ and vacuum wavelength $\lambda = 20 \pi \Delta x$ which is sufficient to make numerical dispersion negligible \cite{KS_Taflove}. The grid is terminated by absorbing boundary conditions with amplitude reflection coefficient estimated at $R \approx 10^{-4}$, what practically means that eventual interference effects are insignificant. The source of radiation is placed in such a way that it passes the middle point of the grid, where the detector is located. After the simulation finishes, the fast Fourier transform is performed on the recorded signal to produce the frequency spectrum. 

The negative index metamaterials are, by necessity, dispersive \cite{Veselago}. To properly simulate wave propagation in such media, the additional Auxiliary Differential Equations (ADE) method \cite{Ziolkowski,Alsunaidi} is used  to calculate the values of material polarization $P$ and magnetization $M$. In the frequency domain, these quantities can be described by a single pole Drude - Lorentz model
\begin{eqnarray}
P(\omega)&=&\frac{f_e}{\omega_{0e}^2 - \omega^2 - i\gamma_e\omega}E(\omega),\nonumber\\
M(\omega)&=&\frac{f_m}{\omega_{0m}^2 - \omega^2 - i\gamma_m\omega}H(\omega),
\end{eqnarray}
where constants $f_e$, $\gamma_e$, $\omega_{0e}$, $f_m$, $\gamma_m$, $\omega_{0m}$ are determined by the metamaterial structure. By shifting to the time domain and using discretization, one obtains
\begin{equation}
P^n = \frac{4 - 2\omega_{0e}^2\Delta t^2}{2 + \gamma_e \Delta t} P^{n-1} + \frac{-2 + \gamma_e\Delta t}{2 + \gamma_e \Delta t} P^{n-2} + \frac{2f_e\Delta t^2}{2 + \gamma_e \Delta t}E^{n-1}
\end{equation}
and analogous equation for $M$. To adapt the algorithm to the simplified Drude model given by Eq. \ref{modDRU}, one can set $f_e = f_m$, $\gamma_e = \gamma_m = 0$ and $\omega_{0e} = \omega_{0m} = 0$.

The infinitely thin, harmonic source of plane waves can be modeled with a good accuracy by imposing source current in a narrow, Gaussian shaped area
\begin{equation}
j_e(x,t)=j_{e0}e^{-(x-x_s)^2/2\sigma^2}sin(\omega t),
\end{equation}
where $x_s$ is the source position and $\sigma$ is the standard deviation describing the width of the peak. The value of $\sigma$ was chosen empirically on the basis of observation that the width of the Gaussian has no significant impact on the frequency $\omega$ and wavelength $\lambda$ of the produced waves as long as $\sigma<<\lambda$. The only factor that affects the frequency of the electromagnetic radiation source is  the time dilation which was included explicitly. Due to the negligible pulse width, the Lorentz contraction was not considered. The amplitude $j_{e0}$ increased to its nominal level in a continuous manner in order to reduce the frequency spectrum of the generated waves and make them quasi-monochromatic.

The FDTD algorithm for EM waves can be adapted to simulation of sound waves. The electric and magnetic fields can be directly transformed into the pressure and velocity \cite{Shu}. Likewise, the permittivity and permeability relate to the dynamic compressibility and density of the medium. Therefore, the propagation of the acoustical waves can be simulated in FDTD code without any modification of the underlying equations. Moreover, the values of parameters used in our simulation can be set to closely replicate the experimental setup described in \cite{Sound_Lee}, facilitating direct comparison of the numerical results to the experimental data.
\section{Solution for sound waves}
\begin{figure}[ht!]
    \centering
    \subfigure[]{\includegraphics[scale=0.22]{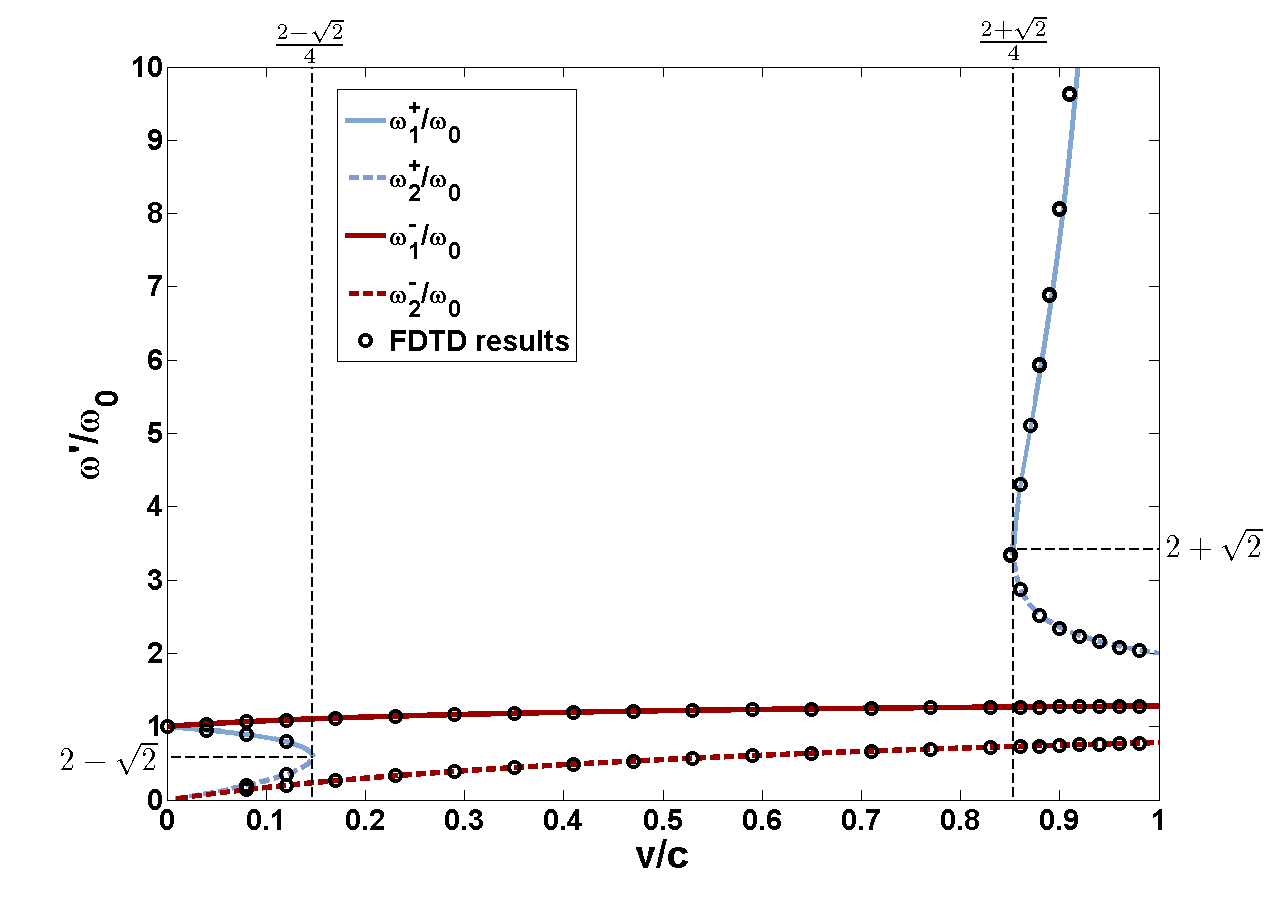}\label{fig:DZ1}}
    \subfigure[]{\includegraphics[scale=0.22]{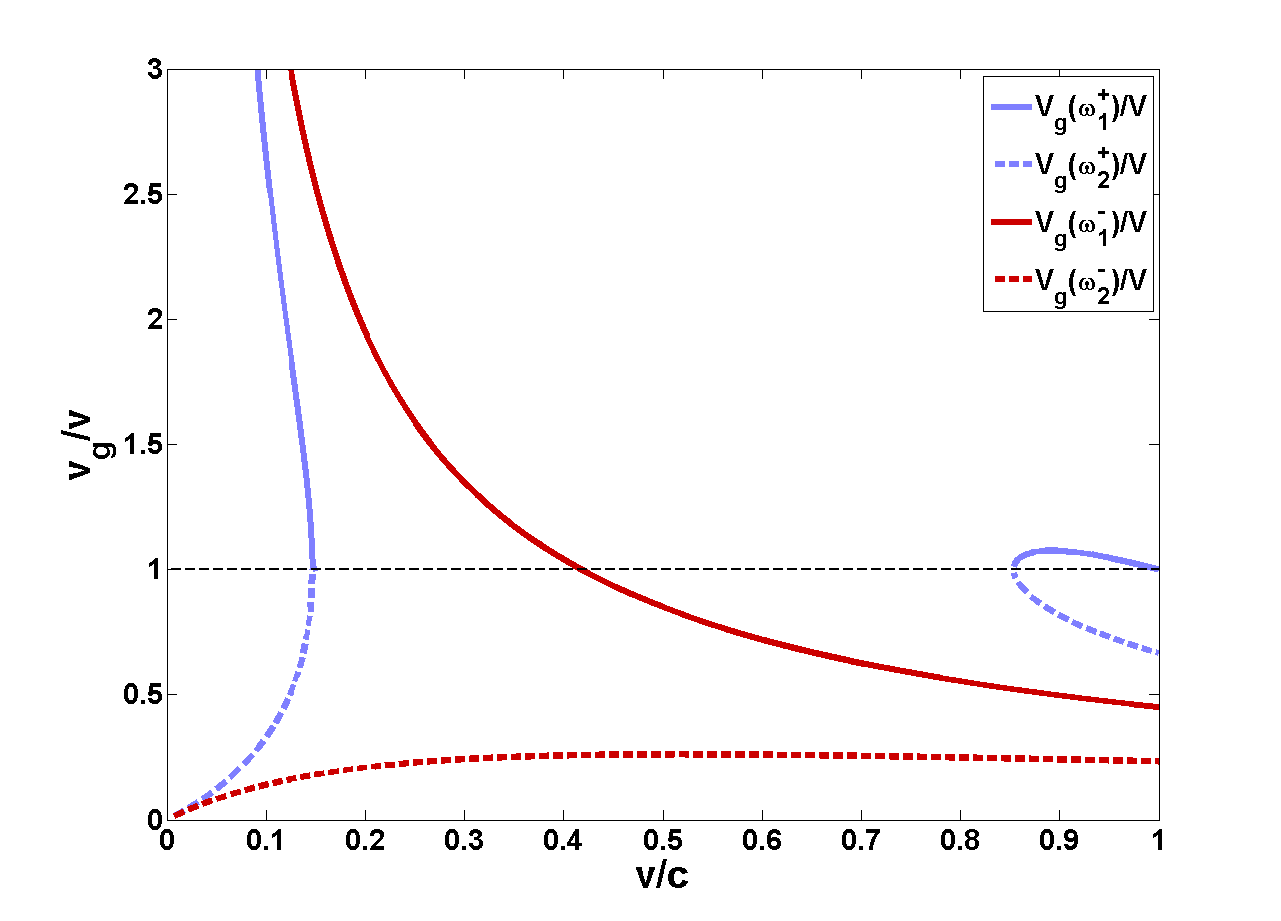}\label{fig:4}}
    \caption{Four acoustic wave frequency components given by Eq. \ref{Final_D} as a function of source velocity for  $n(\omega_0)=-1$ (a). Group velocity of the wave modes as a function of the source velocity (b).}
\end{figure}

\begin{figure}[ht!]
    \centering
    \subfigure[]{\includegraphics[scale=0.19]{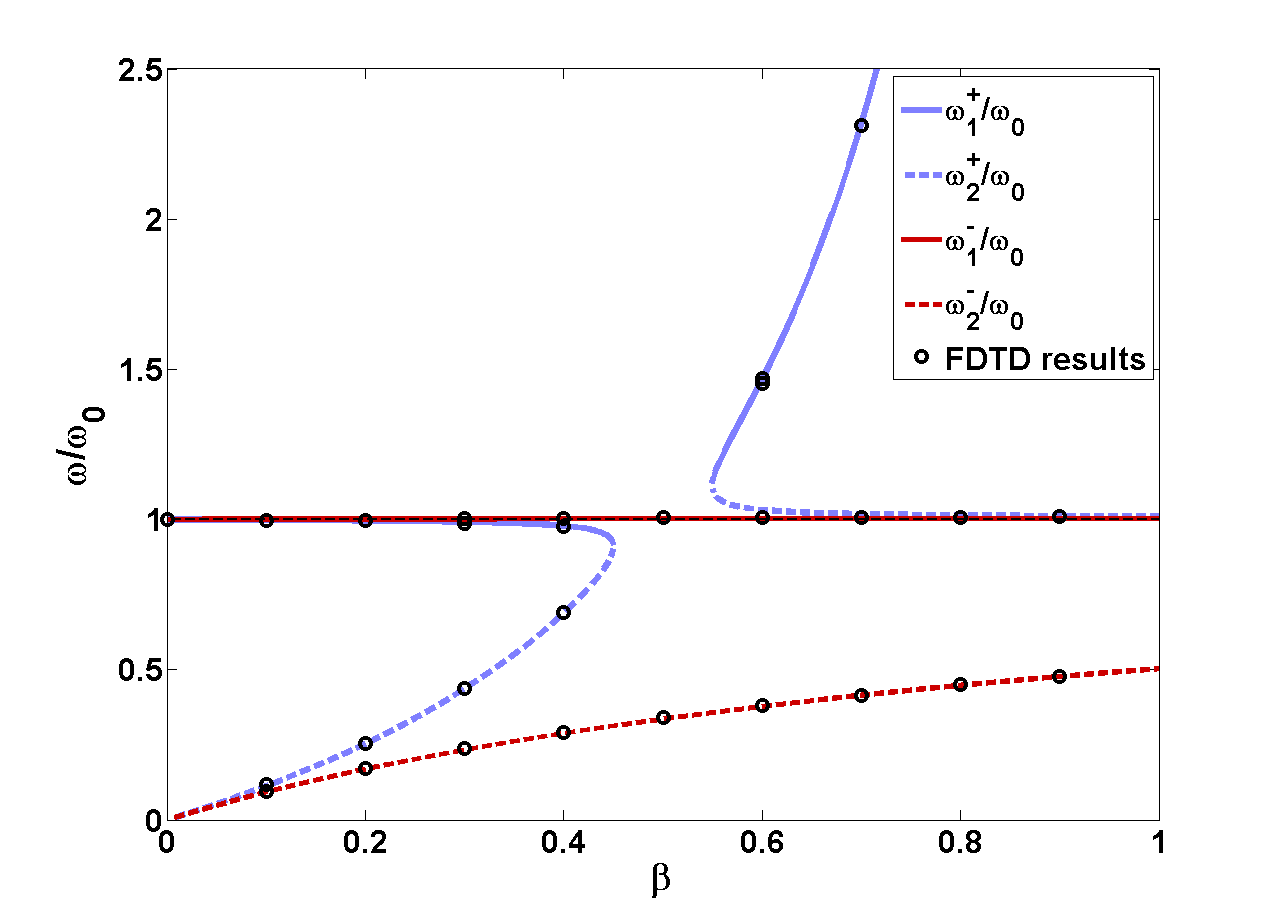}\label{fig:DZ2}}
    \subfigure[]{\includegraphics[scale=0.19]{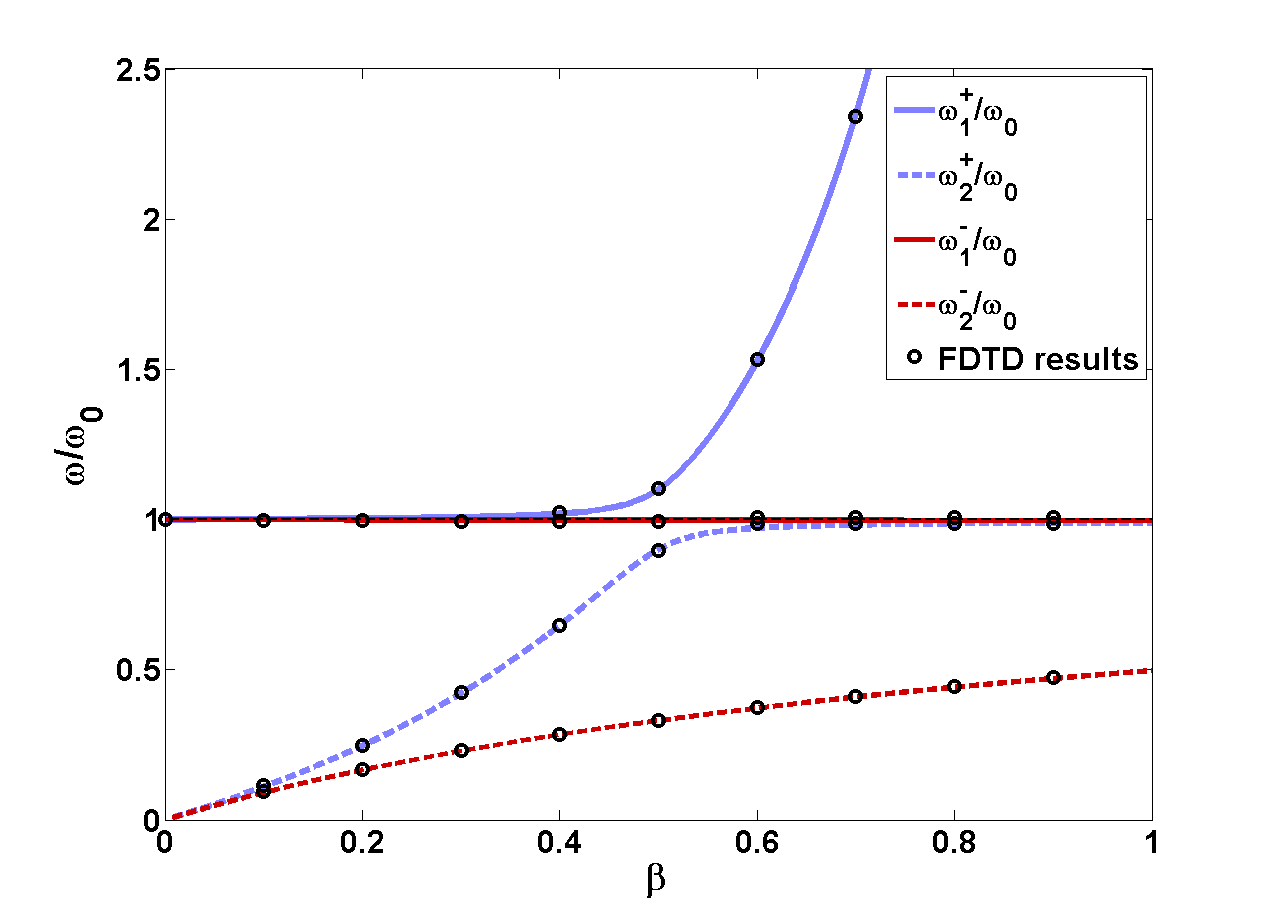}\label{fig:DZ3}}
    \caption{The same as on Fig. \ref{fig:DZ1} but for  $n(\omega_0)=-0.01$ (a) and $n(\omega_0)=0.01$ (b).}
\end{figure}

\begin{figure}[ht!]
    \centering
    \includegraphics[scale=0.18]{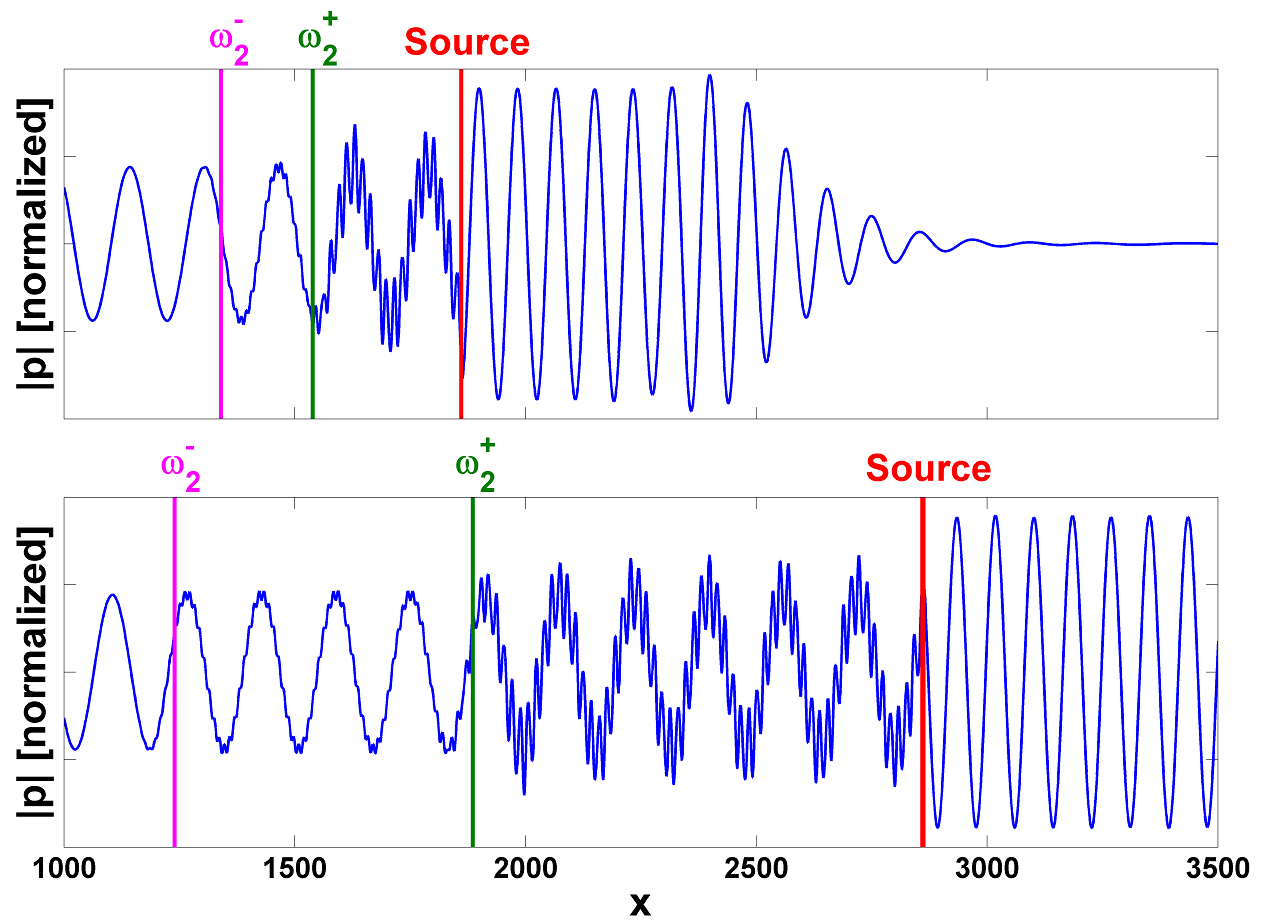}
    \caption{Instantaneous values of the pressure field $p(x)$ for $t=10^4$ (upper figure) and $t=3 \times 10^4$ (lower figure). Approximate boundaries of low frequency modes are marked by vertical lines.}
    \label{fig:5}
\end{figure}

In order to illustrate our theoretical predictions in Eq. \ref{Final_D} for LHM characterized by $n(\omega_0)=-1$,
four Doppler frequencies are plotted on the Fig. \ref{fig:DZ1} along the results of FDTD simulation. Several characteristic points are marked. The obtained frequency values are in excellent agreement with theoretical predictions, with mean relative error $\delta\omega/\omega \approx 10^{-3}$. The propagating waves shown on the figures were observed only in the limited range of source speed where the solutions are real valued. Complex frequency values indicated rapidly decaying waves present only in the near field of the source. The relation between the group velocity of generated frequency modes and the source velocity is shown on the Fig. \ref{fig:4}. One can see that for the additional, forward going wave $\omega_2^+$, the group velocity is always smaller than $V$, so this mode is detectable only behind the source.
\newpage
As expected, at the speed below $V = \frac{2-\sqrt{2}}{4}c \approx 50$ m/s or when the source velocity is higher than $V = \frac{2+\sqrt{2}}{4}c \approx 290~\hbox{m/s}$, four frequency components are detectable. The general form of the solutions shown on the Fig. \ref{fig:DZ1} is maintained as long as $\omega_0^2<\omega_p^2$, so the medium is left-handed. The transition from  LHM to the right-handed medium is presented on the Fig. \ref{fig:DZ2} and \ref{fig:DZ3}, where $n(\omega_0)$ is close to 0 like in the structure considered in \cite{Bongard}. By examining the Eqs.~\ref{Final_D}, one can see that in any negative index medium, there are no real solutions for the $\omega_1^+$ and $\omega_2^+$ modes at $\beta = 0.5$. It should be stressed that for LHM there exists the region of the source velocity where one can detect only waves moving in the direction opposite to the source motion. The range of such velocities increases for decreasing value of~$n$. On the other hand, for any positive value of $n(\omega_0)$, all four modes are present at any $\beta$ (Fig. \ref{fig:DZ3}). Therefore, the Doppler spectrum obtained in the vicinity of $\beta=0.5$ shows a remarkable difference between right- and left-handed media.
In order to better understand the propagation dynamics of these waves, the spatial distribution of the pressure field at two simulation steps was recorded for $\beta = 0.1$ and $n(\omega_0)=-1$. The results are shown on the Fig. \ref{fig:5}. As expected, four distinct wavelengths are visible and the motion of the envelope at the boundary of the area taken by a given mode can be used to estimate its group velocity. The usual modes $\omega_1^+$ and $\omega_1^-$ are propagating at a high speed to the right and to the left respectively. The $\omega_2^+$ mode travels at a slower rate to the right, forming a short trail behind the source. The $\omega_2^-$ mode propagates to the left at a very small group velocity. 
\newpage\noindent
As it was pointed out in \cite{Sound_Lee}, in the case of normal dispersion and negative value of refraction index, the absolute value of $n$ decreases with frequency and, according to Eq. \ref{Lambda}, higher values of $\omega$ correspond to longer wavelength $\lambda$. Therefore, the downshifted wave in front of the source is compressed like in ordinary medium. This effect is clearly visible in spatial field distribution on the Fig. \ref{fig:5}, where the wavelength of the forward going wave is noticeably shorter. For the same reason, the low frequency $\omega_2^\pm$ modes have very short wavelengths. Such a complex effect involving reversed frequency shift and negative refraction index producing unreversed wavelength shift is similar to the case of rotational Doppler effect presented in \cite{Luo}.

\begin{figure}[ht!]
    \centering
    \subfigure[]{\includegraphics[scale=0.18]{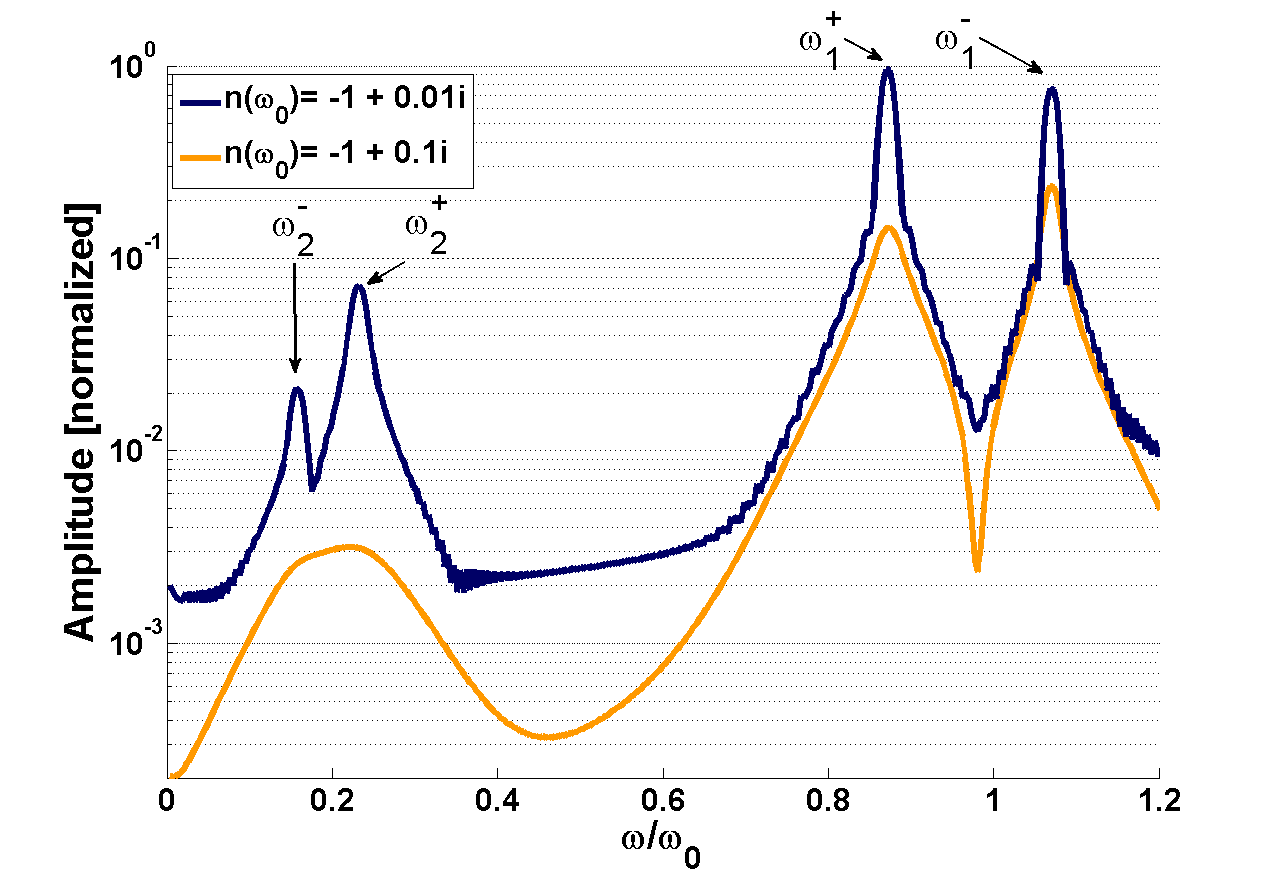}\label{fig:7a}}
    \subfigure[]{\includegraphics[scale=0.18]{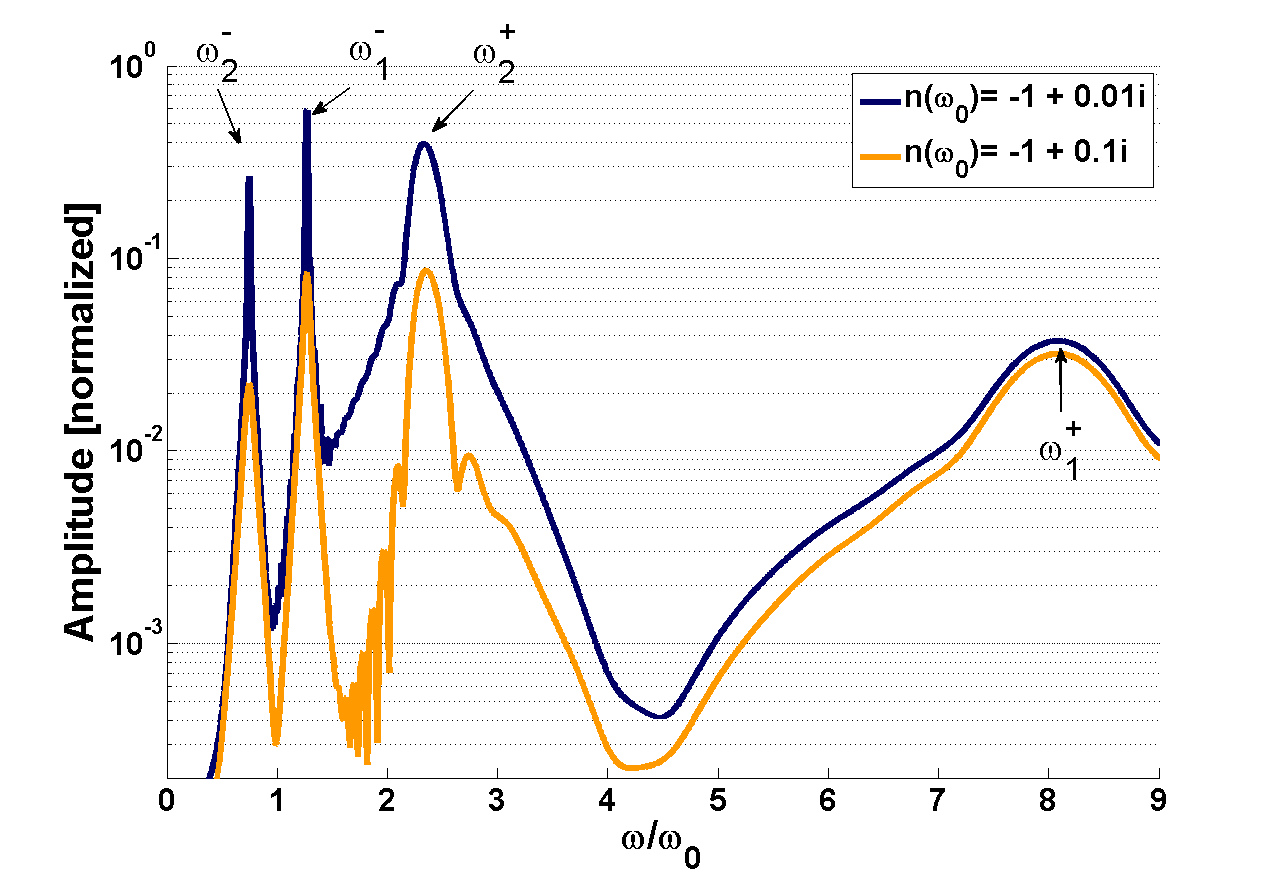}\label{fig:7b}}
    \caption{Frequency spectra for selected values of $n''$ and $\beta=0.09$ (a) or $\beta=0.9$ (b).}
\end{figure}

In order to examine the effect of absorption on the amplitude of Doppler modes, the simulation was performed for a medium characterized by dispersion relation
\begin{equation} \label{lossy_medium}
\widetilde{n}(\omega) = n' + in'' = 1 - \frac{\omega_p^2}{\omega^2 + i \gamma\omega}.
\end{equation} 
The values of $\gamma$ and $\omega_p$ were adjusted to get $n'(\omega_0) = -1$ for various values of $n''$.
The influence of the absorption on the obtained frequency spectra was investigated. The source velocity was set to $V = 0.09 c$ and $V = 0.9 c$, in the regions where four frequency components are present; the results are shown on the Fig. \ref{fig:7a} and Fig. \ref{fig:7b}. In the case of the low source velocity, the amplitude of the additional modes $\omega_2^\pm$ is about order of magnitude lower than the primary ones $\omega_1^\pm$. In the high speed regime, all four modes are comparable. 
\newpage\noindent
As expected, the presence of absorption causes broadening of the frequency peaks due to the reduced time when the signal in the detector has significant amplitude. In the Drude dispersion model used, lower frequency modes are significantly more damped due to the higher value of $n''(\omega)$, making them undetectable at very low source velocity (Fig. \ref{fig:7a}). Additionally, it was observed that the width of the $\omega_1^+$ and $\omega_2^+$ peaks increases as their group velocity approaches the source speed. By moving at a small velocity in relation to the source, those modes spread in a relatively little space around the source in a given time, which leads to a short timeframe where they can be detected. Another detrimental factor is the unavoidable discontinuity of the recorded field at the moment when the source passes the detector.  The general shape of obtained spectra is consistent with theoretical predictions presented in \cite{Lisenkov}. Our theoretical findings agree with the experimental results described by Lee et al  \cite{Sound_Lee}; for $n(\omega_0) \approx -1.7$ and $v= \hbox{5~m/s}$  one obtains two detectable frequencies with ratio  $\omega_1^+/\omega_1^- \approx 1.051$ what is a good match to the reported value $18/17 \approx 1.059$.
\section{Simulation results for electromagnetic waves}
\begin{figure}[ht!]
    \centering
    \includegraphics[scale=0.25]{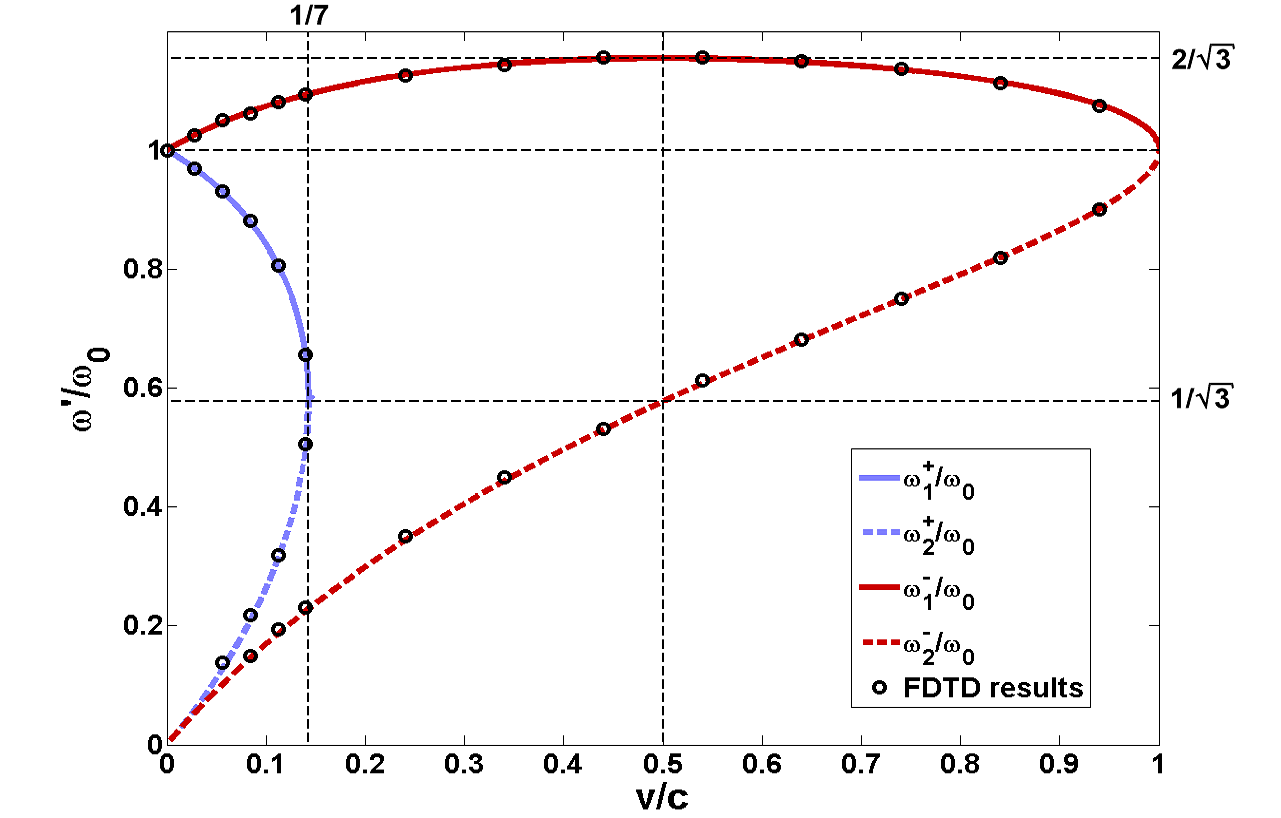}
    \caption{The frequency of four EM waves as a function of source velocity for $n(\omega_0)=-1$.}
    \label{fig:3}
\end{figure}

The solutions for electromagnetic wave modes given by Eq. \ref{Final} were tested in FDTD simulation for LHM characterized by $n(\omega_0)=-1$. The results are shown on the Fig. \ref{fig:3}. When compared to the solution for acoustical waves, there are noticeable  differences at high values of $\beta$, where the relativistic effects play a significant role. Most importantly, the forward going waves are visible only in one area where the source velocity is below c/7. Moreover, there exist absolute limits on the shift magnitude. Minimum possible frequency for usual Doppler mode is $\omega_0/\sqrt{3}$ at $V=c/7$. Maximum frequency is $2\omega_0/\sqrt{3}$ at $V=c/2$. The limit can be understood by examining the implicit Eq. \ref{Dop3}. As the velocity of the source and frequency increase, the refraction index $n(\omega^-) \rightarrow 0$. At some point, nonrelativistic factor $1 + n(\omega^-)\beta$ is dominated by the Lorentz factor $\sqrt{1 - \beta^2}$, causing reduction of frequency, eventually approaching the value of $\omega_0$. This is a significant departure from the solution for nondispersive medium where the asymptotic limits of the two existing frequency components are $0$ and $\infty$.
\section{Conclusion}
The Doppler effect in left-handed medium is both reversed and complex, producing up to four frequency components in the one dimensional case. The presented theoretical predictions and FDTD simulation results for simple Drude model metamaterial are very general in nature and might be treated as an approximation of a wide range of optical and acoustical systems. The existence of the four Doppler modes was confirmed in simulations involving both left- and right-handed media, even with significant absorption. The solutions for sound waves show pronounced difference
of frequency behaviour in these two cases. In LHM, for some range of source velocities dependent on the refracive index $n$, one can observe only waves moving in the direction opposite to the source motion. The numerical source model allowed for smooth, interpolated motion even at a  relatively small speed, yielding results in agreement with the experiment performed by Lee et al \cite{Sound_Lee}. The performed simulations and obtained frequency spectra give insight into the dynamics of the Doppler modes, their properties and dependence on the medium parameters.


\begin{thebibliography}{99}
\bibitem{Veselago}
V. Veselago, \textit{Soviet Physics USPEKHI} \textbf{10}, 509 (1968).

\bibitem{Smith}
D. R. Smith, W. J. Padilla, D. C. Vier, S. C. Nemat-Nasser, and S. Schultz, \textit{Phys. Rev. Lett.} \textbf{84}, 4184 (2000).

\bibitem{Shelby}
R. A. Shelby, D. R. Smith, and S. Schultz, \textit{Science} \textbf{292}, 5514 (2001).

\bibitem{Veselago2}
V. Veselago, L. Braginsky, V. Shklover, and C. Hafner, \textit{J. of Comput. and Theoret.
Nanoscience} \textbf{3}, 1 (2006).

\bibitem{Pendry1}
J. B. Pendry, \textit{Contemporary Physics} \textbf{45}, 191 (2004).

\bibitem{Liu}
Y. Liu and X. Zhang, \textit{Chem. Soc. Rev.} \textbf{40}, 2494–2507 (2011).

\bibitem{GHshift}
P. R. Berman, \textit{Phys. Rev. E} \textbf{66}, 067603 (2002).

\bibitem{Pendry2}
J. B. Pendry, \textit{Phys. Rev. Lett.} {\bf 85}, 3966 (2000).

\bibitem{Dop_Opt}
J. Chen, Y. Wang, B. Jia, T. Geng, X. Li, L. Feng, W. Qian,
B. Liang, X. Zhang, M. Gu, and S. Zhuang, \textit{Nature Photon.} \textbf{5}, 239–245 (2011).

\bibitem{Spinwave}
D. D. Stancil, B. E. Henty, A. G. Cepni, and J. P. Van’t Hof, \textit{Phys. Rev. B} \textbf{74}, 060404(R) (2006).

\bibitem{Seddon}
N. Seddon and T. Bearpark, \textit{Science} \textbf{302}, 1537 (2003).

\bibitem{Sound_Lee}
S. H. Lee, C. M. Park, Y. M. Seo, and C. K. Kim, \textit{Phys. Rev. B} \textbf{81}, 241102 (2010).

\bibitem{Luo}
H. Luo, S. Wen, W. Shu, Z. Tang, Y. Zou, and D. Fan, \textit{Phys. Rev. A} \textbf{78}, 033805 (2008).

\bibitem{Frank}
I. M. Frank, \textit{Izv. AN SSSR, ser. fiz.} \textbf{6}, 3 (1942).

\bibitem{Bazhanova}
A. E. Bazhanova, \textit{Izv. Vyssh. Uchebn. Zaved. Radiofiz.} \textbf{8}, 1100 (1965).

\bibitem{Lisenkov}
I. V. Lisenkov and S. A. Nikitov, \textit{J. Commun. Technol. and Electronics}, \textbf{56}, 6 (2011).

\bibitem{KS_Taflove}
A. Taflove, S. Hagnes, \textit{Computational Electrodynamics: The Finite-Difference Time-Domain Method 2nd ed}, (Artech House, Inc., Norwood, MA, 2000).

\bibitem{Ziolkowski}
R. W. Ziolkowski, E. Heyman, \textit{Phys. Rev. E} \textbf{64}, 056625 (2001).

\bibitem{FDTD_Lee}
J. Y. Lee, J. H. Lee, H. S. Kim, N. W. Kang, and H. K. Jung, \textit{IEEE Trans. on Magn.} \textbf{41}, 5 (2005).

\bibitem{Xiao}
S. Xiao and M. Qiu, \textit{Microwave \& Optical Technology Letters} \textbf{47}, 1 (2005).

\bibitem{Chiou}
Y. P. Chiou, \textit{Reversed Doppler shift in left-handed metamaterials}, IEEE 5th International Conference of Numerical Simulation of Optoelectronic Devices, PD2, Berlin, Germany, September 19-22, (2005).

\bibitem{Wang}
W. Wang, X. Huang, L. Zhou, and C. Chan, \textit{Opt Lett.} \textbf{33}, 369-371 (2008).

\bibitem{Grzegorczyk}
T. M. Grzegorczyk and J. A. Kong, \textit{Phys. Rev. B} \textbf{74}, 033102 (2006).

\bibitem{Berger}
H. Berger, \textit{Am. J. Phys.} \textbf{44}, 851 (1976).

\bibitem{Pendry_8}
J. B. Pendry, A. J. Holden, W. J. Stewart, and I. Youngs, Phys. Rev. Lett. \textbf{76}, 4773 (1996). 

\bibitem{Pendry_6}
J. B. Pendry, A. J. Holden, D. J. Robbins, and W. J. Stewart, \textit{IEEE Trans. Micr. Theory.
Tech.} \textbf{47}, 2075 (1999).

\bibitem{Yee}
K. S. Yee, \textit{IEEE Trans. Antennas Propagat.} \textbf{14}, 302 (1966).

\bibitem{Alsunaidi}
M. A. Alsunaidi, A. Al-Jabr, \textit{IEEE Photonics Tech. Lett.} \textbf{21}, 817 (2009).

\bibitem{Shu}
S. Zhang, \emph{Acoustic metamaterial design and applications} (PhD thesis), University of Illinois at Urbana-Champaign (2010).

\bibitem{Bongard}
F. Bongard, H. Lissek, and J. R. Mosig, \textit{Phys. Rev. B} \textbf{82}, 094306 (2010).

\end{thebibliography}
\end{document}